\documentclass[twocolumn,preprintnumbers,amsmath,amssymb,prl,superscriptaddress]{revtex4-1}

\usepackage[pdftex]{color,graphicx}
\usepackage{latexsym,amsmath,amssymb,gensymb}
\usepackage{hyperref}

\begin{document}
\title{Atomic scale investigation of the volume phase transition in concentrated PNIPAM microgels}

\author{M. Zanatta}\email{marco.zanatta@unitn.it}
\affiliation{Department of Physics, University of Trento, I-38123 Trento, Italy.}

\author{L. Tavagnacco}\email{letizia.tavagnacco@roma1.infn.it}
\affiliation{CNR-ISC and Department of Physics, Sapienza University of Rome, I-00185 Roma, Italy.}

\author{E. Buratti}
\affiliation{CNR-ISC and Department of Physics, Sapienza University of Rome, I-00185 Roma, Italy.}

\author{E. Chiessi}
\affiliation{Department of Chemical Sciences and Technologies, University of Rome Tor Vergata, I-00133 Roma, Italy.}

\author{F. Natali}
\affiliation{CNR-IOM, Operative Group in Grenoble (OGG), c/o Institut Laue Langevin, F-38042 Grenoble, France.}

\author{M. Bertoldo}
\affiliation{Department of Chemical and Pharmaceutical Sciences, University of Ferrara, I-44121 Ferrara, Italy.}
\affiliation{CNR-ISOF, I-40129 Bologna, Italy.}

\author{A. Orecchini}\email{andrea.orecchini@unipg.it}
\affiliation{Department of Physics and Geology, University of Perugia, I-06123, Perugia, Italy.}
\affiliation{CNR-IOM c/o Department of Physics and Geology, University of Perugia, I-06123, Perugia, Italy.}

\author{E. Zaccarelli}\email{emanuela.zaccarelli@cnr.it}
\affiliation{CNR-ISC and Department of Physics, Sapienza University of Rome, I-00185 Roma, Italy.}

\date{\today}

\begin{abstract}
Combining elastic incoherent neutron scattering and differential scanning calorimetry, we investigate the occurrence of the volume phase transition (VPT) in very concentrated PNIPAM microgel suspensions, from a polymer weight fraction of 30~wt\% up to dry conditions. Although samples are arrested at the macroscopic scale, atomic degrees of freedom are equilibrated and can be probed in a reproducible way. A clear signature of the VPT is present as a sharp drop of the mean square displacement of PNIPAM hydrogen atoms obtained by neutron scattering. As a function of concentration, the VPT gets smoother as dry conditions are approached whereas the VPT temperature shows a minimum at about 43~wt\%. This behavior is qualitatively confirmed by calorimetry measurements. Molecular dynamics simulations are employed to complement experimental results and gain further insights into the nature of the VPT, confirming that it involves the formation of an attractive gel state between the microgels. Overall, these results provide evidence that the VPT in PNIPAM-based systems can be detected at different time- and length-scales as well as in overcrowded conditions.
\end{abstract}

\maketitle

\section{Introduction}
\label{sec:intro}
Microgels are colloidal particles composed by macromolecular polymer networks that are dispersed into a solvent. Depending on their specific chemical composition, some microgels can swell or deswell in response to external stimuli~\cite{Fernandez2011}, such as temperature~\cite{WuCPS1994}, ionic strength~\cite{ZhangJACS2005} or cosolvent content~\cite{KremerARCM2019,TavagnaccoJML2019}. This ability of drastically modifying the inner structure and behavior when a slight change is applied in the surrounding environment is the key feature on which smart materials are based. Indeed, the responsive nature of microgels, coupled with their versatility and relatively easy synthesis methods, makes them attractive for several applications, such as drug delivery, biocatalysis, sensing, tissue engineering, to name a few~\cite{OhPPS2008,FernandezACIS2009,KargL2019}. In addition, the possibility of modifying the microgel size \textit{in situ} allows to easily tune its volume fraction $\phi$. In particular, thanks to the soft nature of the particles, $\phi$ can be made very large, well above overlap, enabling to explore ultra-dense, sometimes loosely called jammed, states. Altogether these features paved the way for using microgels as model systems to elucidate fundamental problems in condensed matter physics. Among others, examples are the nucleation of squeezable particles~\cite{ScottiPNAS2016}, frustration in colloidal crystals~\cite{HanNATURE2008}, soft depletion and effective interactions~\cite{BergmanNATCOM2018}, glass~\cite{MattssonNATURE2009} and jamming transitions~\cite{ZhangNATURE2009}, as well as the rheological behavior of such jammed states~\cite{ConleyNATCOM2019}.

Among microgels, the most studied systems are based on poly-(N-isopropyl-acrylamide) (PNIPAM) cross-linked chains~\cite{YunkerRPP2014}. Well-established synthesis protocols allow to selectively produce particles with diameters ranging from about 50~nm~\cite{MeyerMMOL2005,AcciaroL2011} to a few microns~\cite{StillJCIS2013}. The typical internal structure of PNIPAM microgels is characterized by a dense core surrounded by a loose corona of long polymer chains and few cross-linkers~\cite{StiegerJCP2004,ConleySA2017}. At room temperature, water is a good solvent for PNIPAM and is incorporated within the polymer network that swells to a large volume. The complex architecture of this network structure is particularly suitable to confine water molecules and other H-bond liquids. For example, microgels with a PNIPAM concentration $c$ higher than 43~wt\% were found to efficiently prevent water crystallization well below 273~K~\cite{AfrozeJMS2000,VanDurmeMMOL2004}. For such very dense regime, we recently reported an investigation of the dynamical behavior of PNIPAM microgels at low temperatures, well below room temperature. By means of neutron scattering experiments and numerical simulations, it was found that the low-temperature evolution of the mean square displacement (MSD) of the polymer atoms shows a slope change at a temperature $T_C\simeq250$~K, due to the onset of anharmonic motions ~\cite{ZanattaSA2018}. This phenomenon in microgels is similar to the known dynamical transition observed in proteins~\cite{DosterNATURE1989,KatavaPNAS2017} and appears to be strongly driven by PNIPAM-water interactions~\cite{TavagnaccoJPCL2019}.
 
PNIPAM-based microgels are mostly studied because of their thermoresponsive nature~\cite{YunkerRPP2014}. Indeed, when the temperature $T$ is increased, PNIPAM chains undergo a conformational transition toward a globular state: water is partially expelled from the polymer network and the particles collapse~\cite{PeltonJCIS2010}. This process is named volume phase transition (VPT) and, in dilute conditions, it occurs at $T_{VPT}\sim305$~K~\cite{BaeJPSB1990}. Structurally, the VPT appears as a sudden drop of the microgel particle size that is usually detected by means of dynamic light scattering or by X-ray and neutron small angle scattering (e.g.~\cite{SierraMartinACIS2014}). Increasing the polymer concentration to a few weight percent, particle aggregation becomes relevant~\cite{BischofbergerSR2015}. From the dynamical point of view, the VPT of microgels has been studied only in relatively dilute conditions~\cite{SierraMartinACIS2014}, while suspensions of linear polymer chains have been investigated up to a PNIPAM content of about 30 wt\%~\cite{OsakaJCP2009,NiebuurMM2019}. In both cases, a sharp decrease in the MSD measured by neutron scattering experiments was observed. The temperature-polymer volume fraction phase diagram of PNIPAM microgels in highly dilute regime has been successfully described by the Flory-Rehner theory, with the inclusion of a concentration dependent Flory solvency parameter~\cite{FernandezPRE2002} and a similar theoretical approach can represent the effect of hydrostatic pressure over the swelling of microgel particles~\cite{SantosSM2011}. To keep up with the large number of experimental and theoretical results, the role of numerical simulations has recently become more and more crucial for a better understanding of microgels behavior~\cite{RovigattiSM2019}. In particular atomistic simulations have helped evaluating the molecular features that affect the lower critical solution temperature of the polymer and thus the related VPT of microgels~\cite{LonghiCPL2004, Chiessi2010, Alaghemandi2012, Liu2014, Abbot2015, Botan2016, KangJPCB2016, deOlivera2017, Adroher2017, Garcia2018, ConsiglioPCCP2018, MilsterPCCP2019} and they provided useful insights into the molecular mechanism that drives the process~\cite{DeshmukhJPCB2012,TavagnaccoPCCP2018,NetzPCCP2018,PodewitzJPCB2019}. 

In this paper we extend the study of the microgel atomic dynamics across the volume phase transition to the unexplored high-concentration region. By combining elastic incoherent neutron scattering (EINS) and differential scanning calorimetry (DSC) with all-atom molecular dynamic simulations (MD) we study the VPT in PNIPAM microgel suspensions at concentrations ranging from 30 to 95~wt\% (dry). Our findings reveal that the VPT appears in the hydrated suspensions as a sudden drop of the MSD of the polymer atoms onto the value of the dry sample. This highlights a stiffening of the microgel network due to the expulsion of the solvent. The VPT temperature obtained from EINS data, that probes the microscopic dynamics, shows the same concentration dependence as the macroscopic value provided by DSC. Finally, MD simulations carried out on the same space- and time-scale of neutron scattering experiments allow to highlight collective aspects of the VPT in concentrated systems.

\section{Experimental and numerical methods}
\label{sec:exp}
For properly exploiting the capabilities of EINS to single out the dynamics of PNIPAM atoms with respect to those of the solvent, we prepared a set of microgel suspensions of PNIPAM dissolved in D$_2$O. Indeed, in thermal neutron scattering, the cross-section of hydrogen atoms, abundantly and uniformly distributed in the chemical structure of the polymer, is more than one order of magnitude larger than those of other atomic species in our samples, including deuterium. Consequently our EINS experiments mostly probe the atomic dynamics of the PNIPAM polymer, whereas the deuterated solvent contribution remains negligible.

\subsection{Sample preparation}
\label{sec:expSample}
A reliable synthesis protocol was adopted to prepare a set of six PNIPAM microgel powders, hydrated with deuterated water at concentrations ranging from 30 to 95~wt\% (dry).  Microgels were synthesized by precipitation polymerization of N-isopropylacrylamide (NIPAM) in water (0.137~M) in presence of N,N'-methylenebisacrylamide (BIS)(1.84~mM) at $T=343$~K. The reaction was carried out for 10~h in nitrogen atmosphere in presence of 7.80~mM sodium dodecylsulfate as surfactant and potassium persulfate (2.46~mM) as radical initiator. The resulting colloidal dispersion was purified by exhaustive dialysis against pure water, lyophilized and dispersed again in D$_2$O to a final concentration of 10~wt\%, that was determined by thermogravimetric analysis. The particle size of the obtained microgels was characterized by means of dynamic light scattering (Zetasizer Nano S, Malvern). The hydrodynamic diameter was found to be $94\pm3$~nm at 293~K, with a size polydispersity of $0.17\pm0.01$.
Starting from the pristine material, high concentration samples were obtained by evaporating the exceeding D$_2$O in dry atmosphere using a desiccator under moderate vacuum ($\sim10$~mmHg). The concentration was checked by weighting the samples throughout the process. Once reached the target values of 30, 43, 50, 60, 70~wt\%, the samples were sealed in appropriate sample holders and left to homogenize at room temperature for several hours (in the case of few mg samples for DSC) or days (in the case of few g samples for EINS). The dry sample (95~wt\%) was prepared from film casting the PNIPAM dispersion at 10~wt\% up to dryness in Petri dish. The obtained transparent films were milled with an IKA MF 10.1. Cutting-grinding gave rise to a rough powder that was poured and sealed into the sample holder. The dry mass content in the sample was determined by thermogravimetric analysis as mass fraction residua at 430~K. Samples were visually inspected before and after the experiments, appearing homogeneous with no compartmentalization effects and no appreciable morphological changes.

\subsection{Differential scanning calorimetry}
\label{sec:DSC}
Thermal analyses were recorded with a differential scanning calorimeter SII DSC 7020 EXSTAR Seiko equipped with a liquid nitrogen cooling system. The instrument was calibrated with Indium, Zinc and heptane as standards. Small quantities (20-25~mg) of PNIPAM dispersions in D$_2$O at different concentrations were analysed under nitrogen atmosphere (100~mL/min) in hermetic sealed steel pan to keep the concentration constant during the heating protocol. Measurements were carried out at a constant heating rate of 10~K/min.

\subsection{Elastic incoherent neutron scattering}
\label{sec:expEINS}
A neutron scattering experiment on isotropic samples measures the scattered intensity $I(Q,E)$ as a function of exchanged wavevector $Q$ and exchanged energy $E$, which is proportional to the dynamic structure factor $S(Q,E)$ of the sample. The latter is the Fourier transform of the time-correlation function of the density fluctuations of the sample nuclei and thus provides information on both structure and dynamics of the sample atoms~\cite{Lovesey,Bee}. Concerning in particular its energy dependence, $S(Q,E)$ will produce a signal when the neutron exchanged energy $E$ matches one of the typical energies of the sample atomic motions.
As already mentioned, in our PNIPAM-D$_2$O suspensions the dominant contribution to the signal comes from the PNIPAM hydrogen atoms, whose neutron cross section is by far larger than that of deuterium and of the other atomic species in our samples. Moreover, we recall that neutron scattering by nuclei can have both a coherent and an incoherent nature, that bear information about collective motions and single-particle dynamics respectively. In the particular case of hydrogen, the incoherent cross-section ($\sigma_{inc} = 80.27(6)$~b) is largely predominant with respect to the coherent one ($\sigma_{coh} = 1.7583(10)$~b)~\cite{NeutronData}. Therefore, our neutron measurements mostly probe the single-particle dynamics of the PNIPAM hydrogen atoms and thus reflect the average mobility of the polymer network, whereas the solvent dynamics gives negligible contributions.

In the particular case of the EINS technique, the neutron scattering intensity is measured within a narrow energy interval centered at the elastic peak ($E\simeq0$). As such, whenever the sample mobility increases and the atomic movements shift to higher energies, the corresponding scattering signal can eventually move out of the experimental elastic energy window and thus result in an EINS intensity drop. Within the incoherent approximation, which is appropriate to EINS experiments, the measured signal $I(Q,0)$ can be written as~\cite{Bee}
\begin{eqnarray}
	I(Q,0) &=& I_0 \exp\left( -\left\langle \Delta u^2 \right\rangle_{vib} Q^2 \right) \left[ A(Q) \delta(E) \right] \nonumber \\
       &&\otimes R(Q,E),
	\label{Eq:sqw1}
\end{eqnarray}
where $R(Q,E)$ is the experimental resolution function. The Gaussian term is the so-called Debye-Waller factor, which accounts for the Q-dependence of the elastic intensity due to purely harmonic (vibrational) atomic MSD $\left\langle \Delta u^2 \right\rangle_{vib}$. Further atomic mobility due to anharmonic movements, such as relaxational, diffusional or bi-stable dynamical processes, is accounted for by the elastic incoherent structure factor (EISF), $A(Q)$, that further modulates the scattering intensity. The detailed shape of $A(Q)$ depends on the specific nature and spatial geometry of the involved atomic movements. 

EINS measurements were carried out at the high-resolution backscattering spectrometer IN13 of the Institut Laue-Langevin (ILL, Grenoble, France)~\cite{DataDOI}. This instrument has an elastic energy-resolution $\Delta E=8~\mu$eV (FWHM) and covers an interval of exchanged momentum $Q$ from about 0.2 to 4.5~\AA$^{-1}$, thus giving access to motions faster than about 150~ps, occurring in the spatial region between 1 and 30~\AA. We can thus probe the internal dynamics of the microgel, exploring time- and length-scales where polymeric degrees of freedom can be assumed to be in equilibrium also at high PNIPAM concentration.
Samples were measured inside flat Al cells sealed with an In o-ring. The weight of each sample was checked before and after the measurement without observing any appreciable variation. The thickness of each sample was chosen to obtain a nominal transmission of about 90\%, which was then experimentally confirmed by transmission measurements at IN13. Experiments were carried out in the fixed-window elastic mode, thus collecting the $I(Q,0)$, i.e. the elastically scattered intensity as a function of $Q$. Data were corrected to take into account for incident flux, cell scattering and self-shielding. The $I(Q,0)$ of each sample was normalized with respect to a vanadium standard to account for detector efficiency fluctuations. 

\subsection{All-atom molecular dynamics simulations}
\label{sec:expMD}
All-atom molecular dynamics (MD) simulations were performed at a PNIPAM concentration of 30~wt\%. We considered two different molecular models. The first one is based on a system that we recently developed~\cite{ZanattaSA2018,TavagnaccoJPCL2019} which describes the microgel as a polymer network composed by 6 interconnected PNIPAM 30-mers, i.e. oligomers formed by 30 repeating units. This system represents an all-atom model of a portion of the microgel particle, and, thanks to extra-boundaries connectivity between the chains, it mimics the percolation of the whole polymer network. However, the infinite covalent connectivity of this model may affect the polymer behavior and could be  unsuitable for properly observing a volume phase transition. To overcome this problem, we also considered an aqueous suspension of linear polymer chains, each composed of 30 repeating units. In both models,  chains are described as atactic stereoisomers by assuming a Bernoullian distribution of meso and racemo dyads, with a content of racemo dyads of 55\%. Amide groups belonging to the side chains are represented with a trans arrangement.
Simulations were performed with the GROMACS 2016.1 software\cite{Pall2015,AbrahamSX2015}. PNIPAM is modeled using the All-Atom Optimized Potentials for Liquid Simulations (OPLS-AA) force field~\cite{JorgensenJACS1996} with the implementation by Siu et al.~\cite{SiuJCTC2012}, which has been shown to properly reproduce the lower critical solution temperature behavior of a linear chain at infinite dilution~\cite{ChiessiJPCB2016,TavagnaccoPCCP2018}. Water is treated with the Tip4p/2005 model~\cite{AbascalJCP2005} that is known to correctly describe the experimental temperature dependence of several properties of liquid water, including dynamical properties, in the investigated range of temperatures.
The network model was simulated in a $T$-range between 283~K and 323~K, with a temperature step of 10~K. At each temperature, the system was equilibrated in a pressure bath maintained by the Parrinello-Rahman barostat~\cite{ParrinelloJAP1981,NoseMP1983} up to a constant density value, i.e. tot-drift lower than $2 \times 10^{-3}$~g~cm$^{-3}$ over 20~ns. Trajectory data were then collected for 330~ns in the NVT ensemble.
MD simulations of suspensions of polymer chains were carried out on two systems composed by 16 and 48 chains, respectively. The starting configurations were originated by randomly distributing in a cubic box the chains with a conformation taken from a single-chain trajectory equilibrated at 283 K\cite{TavagnaccoPCCP2018}. The system was hydrated and energy was minimized with a tolerance of 100 kJ~mol$^{-1}$nm$^{-1}$. Equilibration was carried out in a pressure bath controlled by the Berendsen barostat~\cite{BerendsenJCP1984} and simulations were then performed in the NPT ensemble by using the Parrinello-Rahman barostat~\cite{ParrinelloJAP1981,NoseMP1983}. To avoid any influence of the system history, simulations were started from the same initial configuration at each temperature. The suspension of 16 chains was investigated at 7 temperatures (283~K, 293~K, 303~K, 308~K, 313~K, 318~K and 323~K) whereas that composed by 48 chains was studied at 4 temperatures (283~K, 293~K, 303~K and 313~K). Additional simulations of 48 linear chains were performed with the previously described procedure but applying a restraint to the position of backbone tertiary carbon atoms. In this case, PNIPAM chains were equilibrated in vacuum at 313~K with the Berendsen barostat and simulated for 100~ns in the NVT ensemble. The resulting aggregated configuration was hydrated and NPT simulations were carried out at 308 and 313~K. Trajectory data were collected for $\sim 0.45~\mu$s or for the restrained simulations at $T > T_{VPT}$ for $\sim 0.25~\mu$s.
For all the investigated systems, the leapfrog integration algorithm was employed with a time step of 2~fs~\cite{HockneyMCP1970}. The length of bonds involving hydrogen atoms was kept fixed with the LINCS algorithm~\cite{HessJCC1997}. The Parrinello-Rahman and the Berendsen barostats were used to maintain pressure at 1~bar with a time constant of 2~ps and 1~ps, respectively. Cubic periodic boundary conditions and minimum image convention were applied. Temperature was controlled with the velocity rescaling thermostat coupling algorithm with a time constant of 0.1~ps~\cite{BussiJCP2007}. Electrostatic interactions were treated with the smooth particle-mesh Ewald method with a cutoff of non-bonded interactions of 1~nm~\cite{EssmannJCP1995}. The trajectory sampling was set to 0.2~frame/ps, while last 100~ns were typically considered for data analysis.

\section{Results}
\label{sec:Res}

\subsection{Estimates of the Volume Phase Transition temperature}
\label{sec:resVPTT}
\begin{figure}[htbp]
	\centering
	\includegraphics[width=0.4\textwidth]{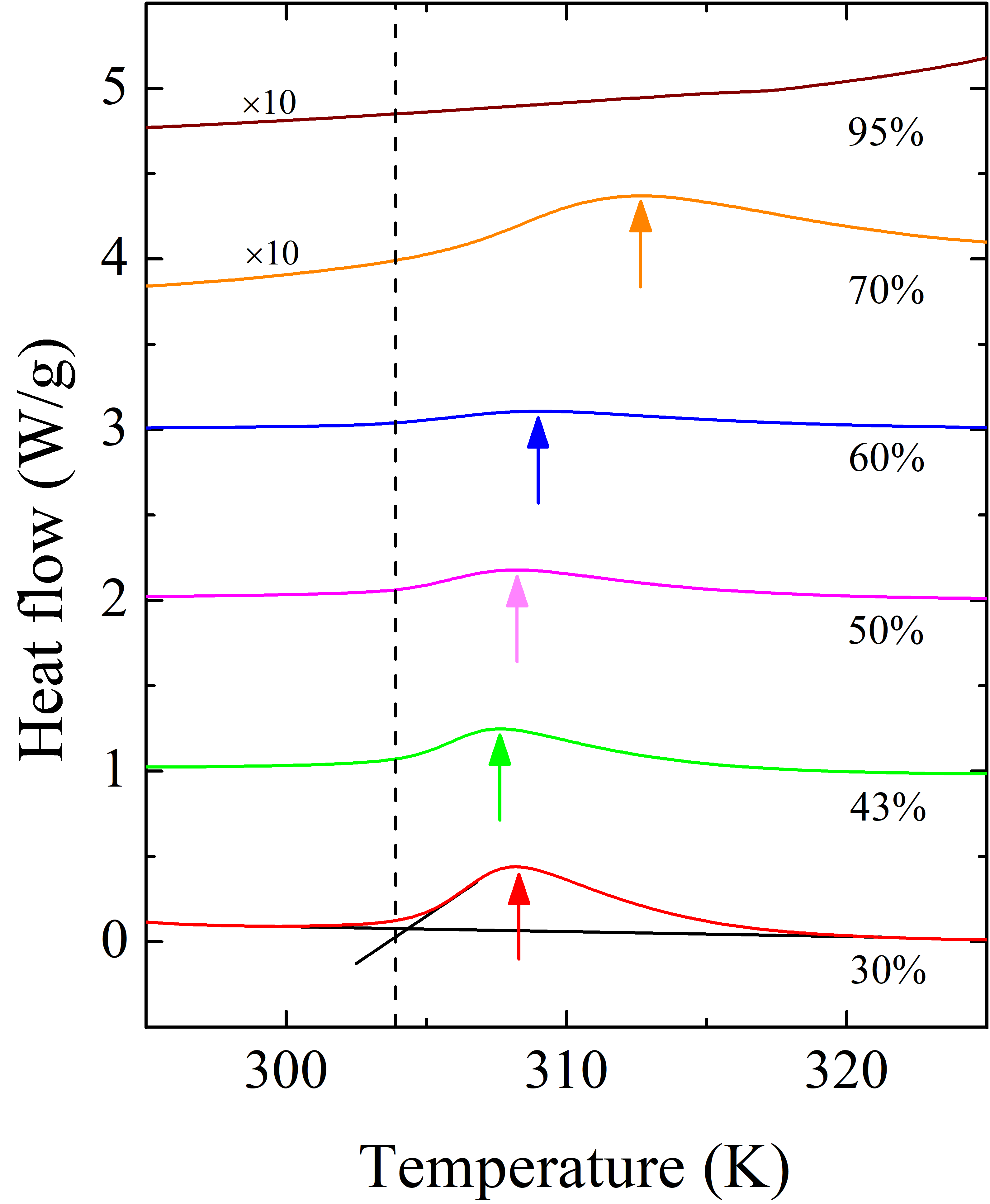}    
	\caption{Thermograms recorded for PNIPAM microgels in D$_2$O as a function of polymer mass fraction. The heat flow is normalized to the mass of the measured sample. In order to better appreciate the evolution of the peak intensity, a constant baseline was subtracted to each curve and data were then shifted along the vertical axis. The thermograms for $c\geq70$~wt\% are magnified by a factor 10. The position of $T_{VPT}$ is marked by an arrow whereas the black dashed line represents the average of the $T_{ONS}$ values. The definition of $T_{ONS}$ is shown for the 30 wt\% sample (black solid lines).}
	\label{fig1}
\end{figure}

Figure~\ref{fig1} shows the measured thermograms for the investigated samples. The volume phase transition appears as an endothermic peak in the region between 300 and 318~K. This peak can be described by two characteristic temperature values: the onset temperature $T_{ONS}$ and the VPT temperature $T_{VPT}$. The former is defined as the intersection point of the extrapolated baseline and the inflectional tangent at the beginning of the peak as shown in Fig. \ref{fig1} for the 30~wt\% sample (solid black lines). This temperature is found to be  almost constant for all studied PNIPAM concentrations ($T_{ONS}\sim304$~K). Conversely, $T_{VPT}$ is identified by the maximum of the endothermic peak (marked by an arrow in Fig.~\ref{fig1}). As a function of PNIPAM concentration, $T_{VPT}$ decreases at first, reaches a minimum at $\sim$43~wt\% and then grows again for higher concentrations. Moreover, on increasing the polymer content, the transition peak progressively smoothens and finally vanishes when dry conditions are approached. This indicates that the enthalpy variation associated to the process, which is proportional to the peak area, decreases and goes to zero at PNIPAM concentrations roughly above 70~wt\%. Overall, the behavior qualitatively reproduces that observed for PNIPAM linear chains in water~\cite{AfrozeJMS2000,VanDurmeMMOL2004}. Finally, we note that the use of D$_2$O as a solvent shifts $T_{VPT}$ of about 3~K upwards as compared to microgels suspensions in H$_2$O~\cite{BerndtAC2006}.

\begin{figure*}[htbp]
	\centering
  \includegraphics[width=\textwidth]{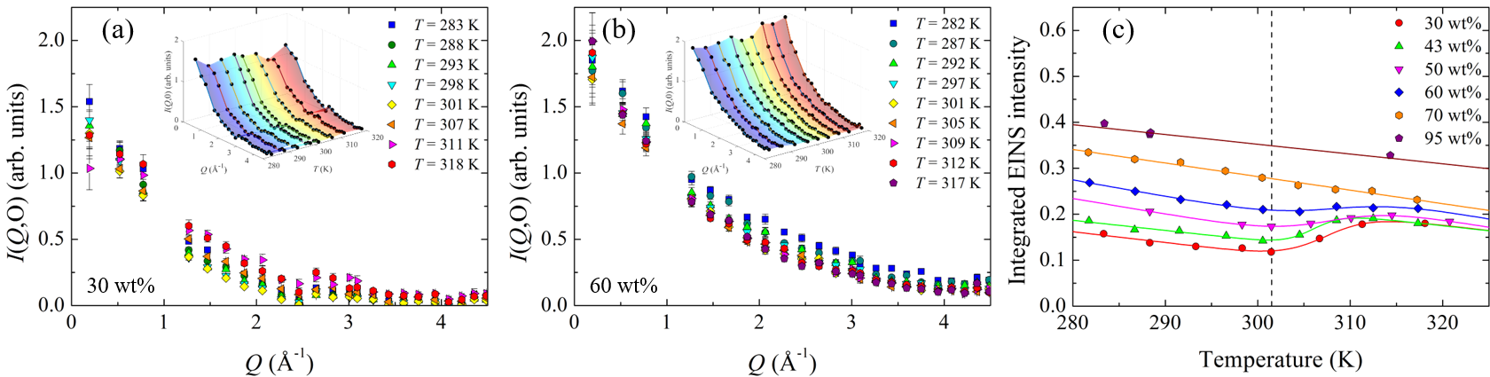}    
  \caption{Temperature evolution of the elastic incoherent intensity $I(Q,0)$ for two samples at different PNIPAM concentrations: (a) 30~wt\% and (b) 60~wt\%. (c) Temperature evolution of the integrated elastic intensity for all measured samples. Data are normalized to 1 when $T$ goes to 0 using the results shown in Ref.~\cite{ZanattaSA2018} except for 30~wt\% sample which is normalized to the 43~wt\% intensity above $T_{VPT}$. The solid lines represent a fit to the data obtained with Eq~\ref{Eq:VPT}. The dry sample is fitted with a straight line. The dashed line shows the onset of the transition, see text.}
	\label{fig2}
 \end{figure*}

Figures~\ref{fig2}(a) and (b) show the temperature evolution of  $I(Q,0)$ versus $Q$ for the 30 wt\% and 60 wt\% samples, respectively. In order to compare the $T$-behavior of the samples, Fig.~\ref{fig2}(c) displays the integrated elastic intensity, i.e. the $Q$-integral of $I(Q,0)$ for all the investigated PNIPAM contents. This provides a first qualitative insight on the dynamical behaviour of the system. For $c\geq43$~wt\%, data were normalized to 1 at $T=0$~K using the low-temperature measurements reported in Ref.~\cite{ZanattaSA2018}. Upon heating, the integrated elastic intensity of the hydrated samples initially decreases up to $T\sim$ 300~K in a linear fashion, and then exhibits a sudden increase at the VPT temperature. The latter is not visible for the dry sample, confirming the calorimetry data.

\begin{figure}[htbp]
	\centering
	\includegraphics[width=0.4\textwidth]{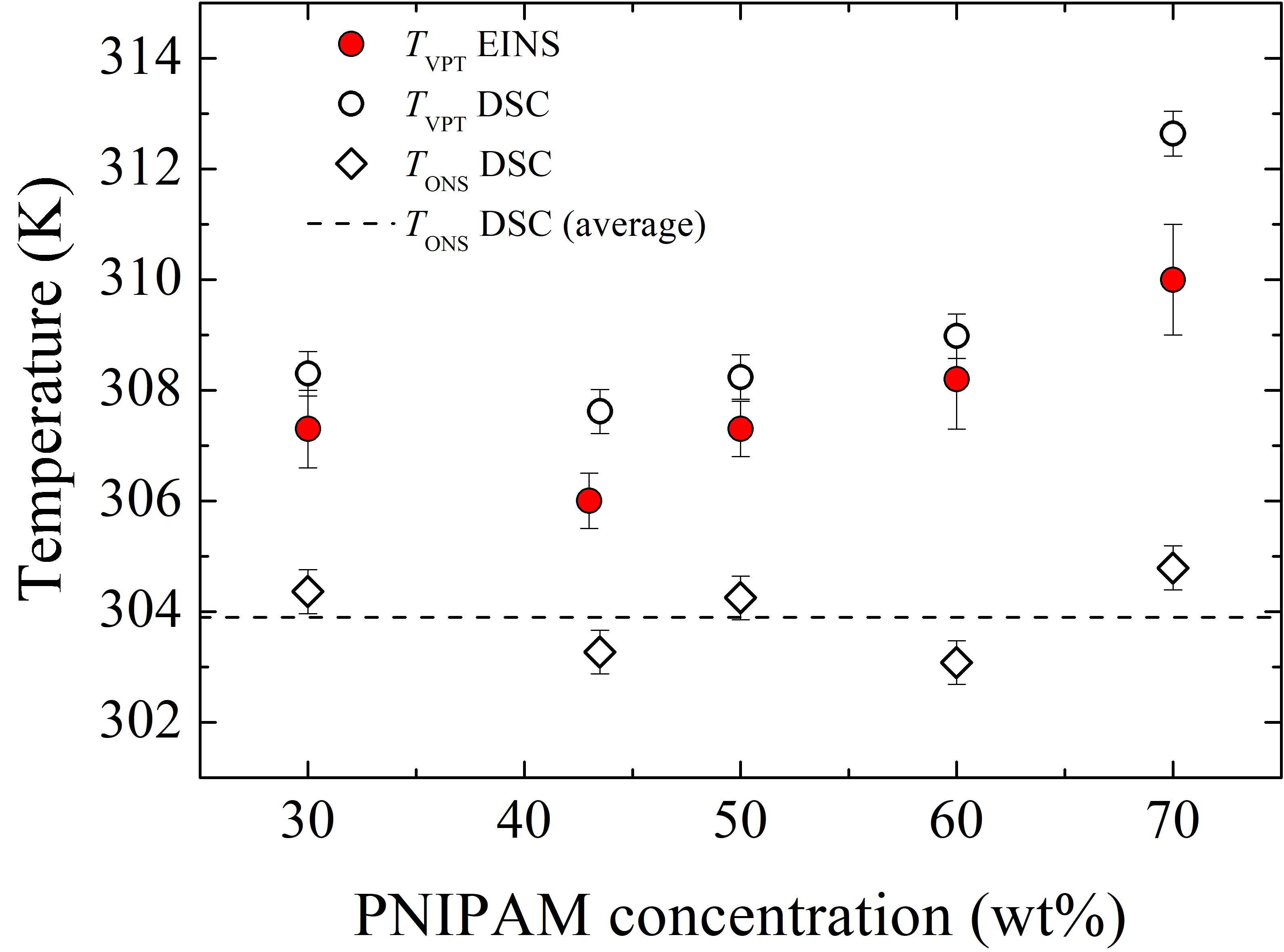}    
  \caption{Comparison between the concentration dependence of the VPT temperature measured by means of DSC (black open circles) and that obtained by EINS experiments (red solid circles). The DSC onset temperature (black open diamonds) is also reported. The black dashed line represents the average of the onset values measured at all the studied concentrations.}
	\label{fig3}
\end{figure}

It is interesting to note that for $T<T_{VPT}$ samples are characterized by a different amplitude of the integrated elastic intensity, whereas they tend to a roughly constant value for $T>T_{VPT}$. Exploiting this behavior, the 30~wt\% sample was normalized to the 43~wt\% data for  $T>T_{VPT}$, since low-temperature measurements are prevented by water crystallization in this sample. While at low $T$ all data follow almost the same linear slope, a deviation occurs at a temperature that seems independent of PNIPAM concentration (black dashed line in Figure~\ref{fig2}(c)) in a way that recalls the behavior of the onset temperature in the calorimetry data. Furthermore, the water content determines the amplitude of the transition that becomes barely visible for the 70~wt\% sample.
The shape of the integrated elastic intensity data can be modeled by some critical function, which would reproduce the occurrence of a transition and allow a quantitative evaluation of the transition temperature. Among the possible function tested, we have adopted a hyperbolic tangent superimposed to a linear background, which provides robust fits to the data as shown in Figure \ref{fig2}(c). Its equation reads
\begin{eqnarray}
	f(T)=mT+q+A\tanh(k(T-T_{VPT})),
	\label{Eq:VPT}
\end{eqnarray}
where $m$ and $q$ are  the slope and the intercept of the linear behavior below $T_{VPT}$, while $A$ and $k$ control amplitude and width of the transition, respectively. From the fits to the data, we can thus extract an estimate of the volume phase transition temperature, that is reported as a function of PNIPAM concentration in Fig.~\ref{fig3}. This quantity can be considered as a microscopic estimate of $T_{VPT}$, as opposed to the macroscopic transition temperature measured by DSC experiments. A comparison between the two is shown in Fig.~\ref{fig3}, clearly indicating that the two experimental techniques, although working at very different time- and length-scales, still detect the same concentration dependence, with a shift of the macroscopic $T_{VPT}$ with respect to the microscopic one by roughly 1~K. As a function of PNIPAM concentration, $T_{VPT}$ decreases slightly between 30~wt\% and 43~wt\% and then grows monotonically. Interestingly, the minimum value of $T_{VPT}$ is found at $c\sim 43$~wt\%, a concentration that also represents the minimum polymer content necessary to avoid water crystallization below 273~K, as determined in Ref.\cite{ZanattaSA2018}.

\subsection{Temperature and concentration dependence of the PNIPAM atomic dynamics}
\label{sec:resMSD}
\begin{figure}[htbp]
	\centering
  \includegraphics[width=0.4\textwidth]{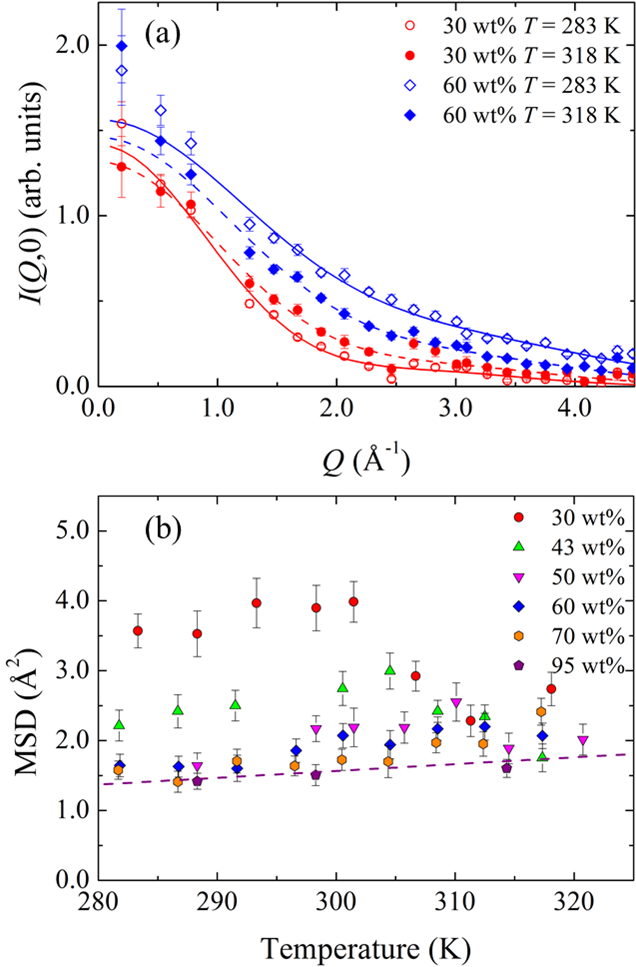}    
  \caption{(a) Typical fits of the $I(Q,0)$ using the double well model for PNIPAM 30\% and 60\% before (283~K) and after (318~K) the VPT; (b) MSD calculated using the double-well model for the measured samples.}
  \label{fig4}
\end{figure}

To get more detailed and quantitative information on the polymer dynamics across the VPT, the measured $I(Q,0)$ has to be fitted with an appropriate model for the EISF $A(Q)$. A widely employed model to fit EINS data of complex polymeric systems is the well-known double-well model~\cite{DosterNATURE1989,KatavaPNAS2017}, that we have already successfully applied to PNIPAM microgel samples~\cite{ZanattaSA2018}. Within this model, all hydrogen atoms are supposed to be dynamically equivalent and may jump between two sites characterised by different potential energy wells located at a distance $d$. The EISF takes then the form:
\begin{eqnarray}
	A(Q) =\left[1 - 2 p_1 p_2 \left(1 - \frac{\sin(Qd)}{Qd} \right) \right],
	\label{Eq:eisf}
\end{eqnarray}
where $p_1$ and $p_2$ are the probabilities of finding the hydrogen atom in the first or second potential well. With this specific assumption, the elastic intensity of Eq.~\ref{Eq:sqw1} now reads:
\begin{eqnarray}
	I(Q,0)&=&I_0\exp\left(- \left\langle \Delta u^2 \right\rangle_{vib} Q^2 \right)\nonumber\\
        &&\times\left[1-2p_1p_2\left(1-\frac{\sin(Qd)}{Qd}\right)\right],
	\label{Eq:doublewell}
\end{eqnarray}
where $I_0$ is a scale parameter and $\left\langle\Delta u^2\right\rangle_{vib}$ corresponds to the already mentioned harmonic vibrational MSD. Typical examples of the fits to the data with Eq.~\ref{Eq:doublewell} are shown in Figure~\ref{fig4}(a) for 30 and 60~wt\% samples. Within this model, where a transition between the two states represents a jump between conformational substates in the free energy surface, the amplitude of the total 3-dimensional MSD (harmonic plus anharmonic contributions) is given by the relationship~\cite{KatavaPNAS2017}:
\begin{eqnarray}
	\mbox{MSD}&=&-6\left(\frac{d\ln I(Q)}{dQ^2}\right)_{Q=0}\nonumber\\
            &=&6\left\langle\Delta u^2\right\rangle_{vib}+2p_1p_2 d^2.
	\label{Eq:MSD}
\end{eqnarray}
With this analysis, we obtain the MSDs as a function of temperature for all investigated PNIPAM concentrations, as reported in Figure~\ref{fig4}(b). For all hydrated samples, the MSDs increase linearly up to the VPT and then decrease suddenly when the microgel collapses, tending to the limiting value measured for the dry system. In the latter case, the MSD increases linearly at all temperatures.

Altogether, EINS data reveal that the VPT manifests itself also at the atomic scale, resulting in a stiffening of the PNIPAM network, whose dynamics for $T>T_{VPT}$ approaches that of a dry sample, possibly with a residual solvation of the microgel~\cite{WuPRL1998,FernandezPRE2002} as shown in the following.

\begin{figure}[htbp]
	\centering
  \includegraphics[width=0.4\textwidth]{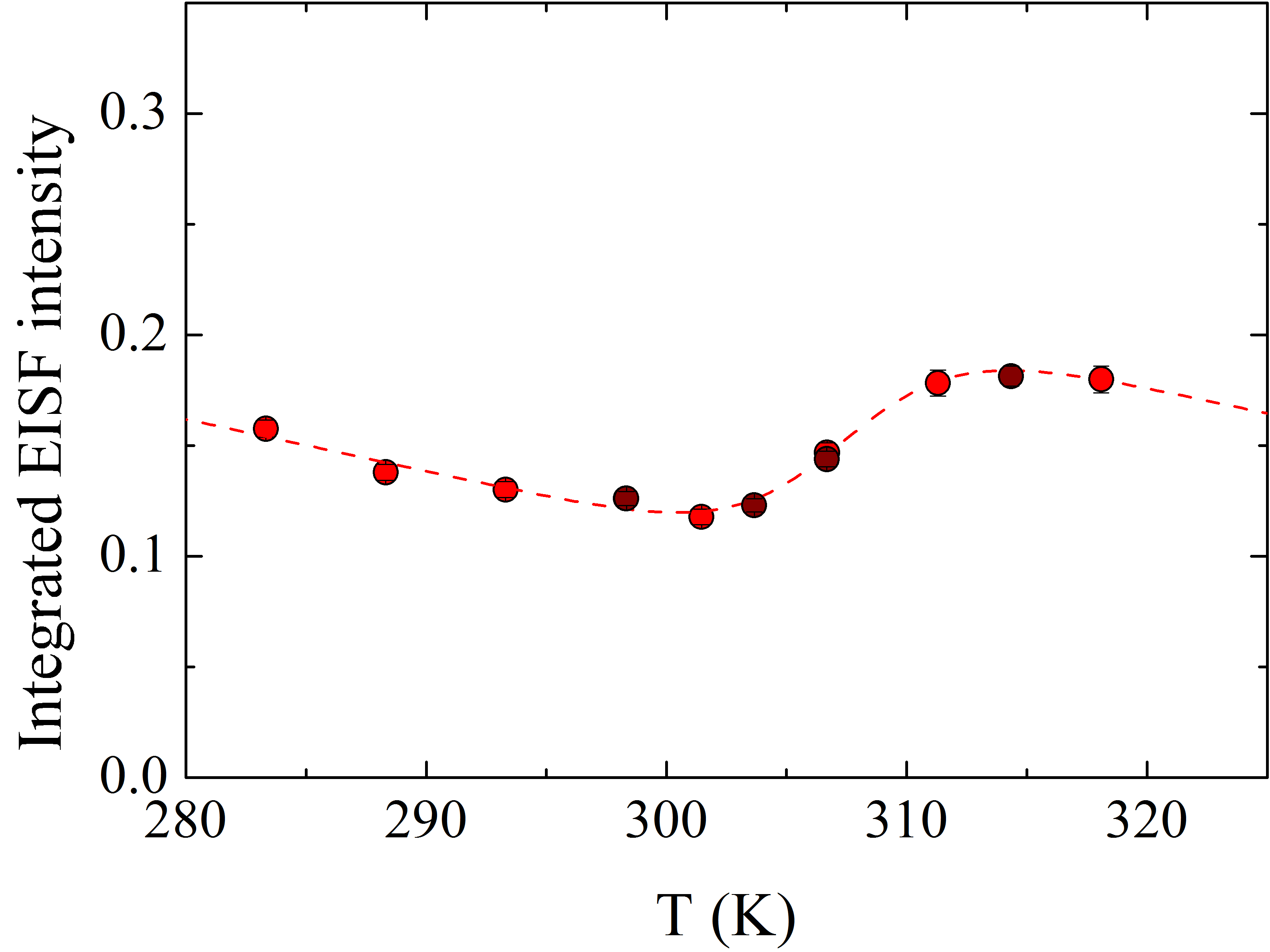}    
  \caption{Integrated elastic intensity measured for PNIPAM 30 wt\% in two subsequent experimental runs. The first measurement set (red circles) was acquired by heating the sample from room temperature to 318~K, whereas the second one (brown circles) was repeated on the same sample after keeping it for 3 days at room temperature. The dashed line is the critical fit by Eq.~\ref{Eq:VPT}, also shown in Fig.~\ref{fig2}.}
  \label{fig5}
\end{figure}

Finally, we tested the reversibility of the VPT at the atomic level by repeating the EINS measurements on the 30~wt\% sample. The first set was acquired by heating from 283~K to 318~K. The phase-separated sample was cooled and kept at room temperature for three days and then measured again from 298~K to 315~K. Figure \ref{fig5} shows the comparison of the integrated elastic intensities, normalized to the same value at 298~K, as obtained from the two sets of measurements. The temperature evolution of the data is found to be independent of thermal history, thus confirming the complete reversibility of the swelling-deswelling process at the atomic level even at high concentrations.

\subsection{Reproducing the VPT in simulations of concentrated PNIPAM suspensions}
\label{sec:resSim}
\begin{figure}[htbp]
	\centering
  \includegraphics[width=0.4\textwidth]{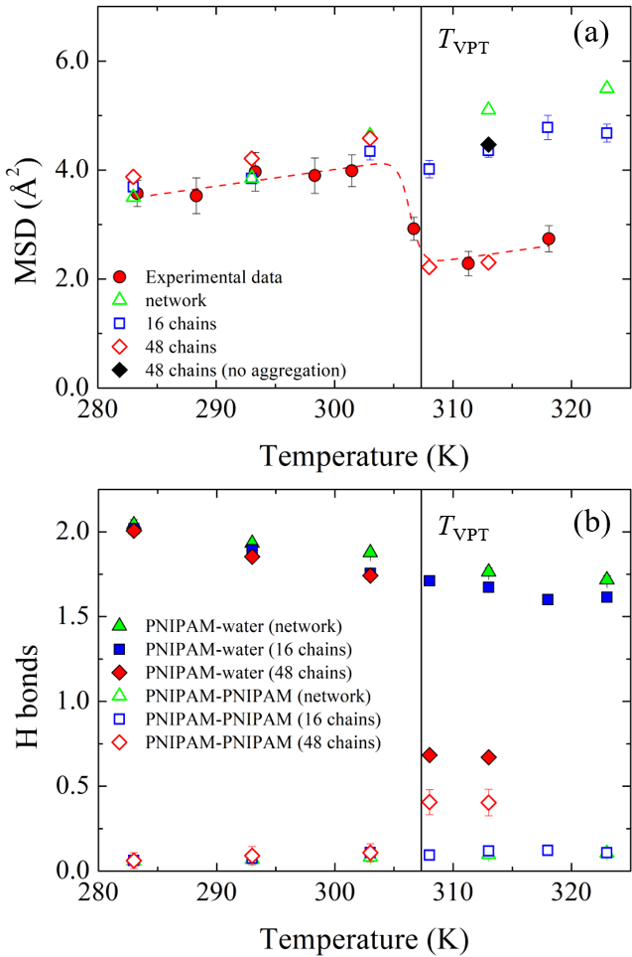}    
  \caption{(a) Comparison between the experimental MSD of the PNIPAM 30~wt\% sample (red circles) and the numerical MSDs calculated for the different modelled systems: polymer network (green open triangles), suspension of 16 linear chains (blue open squares), and suspension of 48 linear chains (red open diamonds). The black diamond represents the MSD obtained for the suspension of 48 linear chains without explicitly considering the polymer aggregation during the collapse. (b) Temperature evolution of PNIPAM-PNIPAM H-bonds numbers for the suspension of 16 linear chains (blue open squares), 48 linear chains (red open diamonds) and polymer network (green open triangles) and PNIPAM-water H-bonds numbers for the suspension of 16 linear chains (blue squares), 48 linear chains (red diamonds) and polymer network (green triangles). Values are normalized to number of PNIPAM residues. The experimental value of $T_{VPT}$ obtained by DSC is marked with a black dashed line in both panels.}
	\label{fig6}
\end{figure}

We now compare how the dynamical properties of different computational models of microgels change across the transition to gain insights into the molecular mechanism of the VPT. We probed the PNIPAM atomic dynamics by calculating the MSD of the polymer hydrogen atoms as a function of time and by looking at the temperature dependence of its value at 150~ps, that coincides with the experimental resolution, as reported in Fig.~\ref{fig6}(a). This procedure was shown to be suitable to allow a direct comparison between numerically and experimentally determined MSD at low temperatures~\cite{ZanattaSA2018}. The MSD of PNIPAM hydrogen atoms was calculated directly from the trajectory using the following equation:
\begin{eqnarray}
	\mbox{MSD}(t)=\langle |\textbf{r}_p(t)-\textbf{r}_p(0)|^2 \rangle,
	\label{Eq:MSDsim}
\end{eqnarray}
where $\textbf{r}_p(t)$ and $\textbf{r}_p(0)$ are the position vector of the PNIPAM hydrogen atom at time $t$ and $0$, with an average performed over both time origins and hydrogen atoms. We performed these calculations for the different models described in the Methods. In Figure~\ref{fig6}(a) we compare the numerical results for both  network model and linear chains with the experimentally determined MSDs.

We find that, for $T<T_{VPT}$, the MSDs linearly increase with temperature for all simulated systems, in quantitative agreement with the experimental values without any scaling factor. However, for $T>T_{VPT}$, some of the models show clear deviations with respect to experiments. In particular, while the MSD value obtained from EINS measurements exhibits an abrupt reduction by roughly a factor 2 close to the VPT, the MSD of the polymer network linearly increases in the whole investigated temperature range. On the other hand, the MSD of the suspension of 16 linear chains  only shows a weak contraction at the VPT. Similar behavior is found for 48 linear chains. In order to reproduce the experimental behavior, we then enhance the aggregation of the system by including positional restraints, as described in Methods, for the suspension of 48 linear chains. For this last model, the experimental dynamical behavior is recovered also in the simulations, without any adjustable parameter. These findings indicate that collective aggregation processes strongly influence the dynamics for $T>T_{VPT}$.

We further monitored local structural changes in the polymer matrix by studying the temperature dependence of hydrogen bonding interactions. PNIPAM-PNIPAM and PNIPAM-water hydrogen bonds were defined through the geometric criteria of a donor-acceptor distance $(\mbox{D-H} \cdots \mbox{A})$ lower than 0.35~nm and an angle $\theta(\mbox{D-H} \cdots \mbox{A})$ lower than 30$^{\circ}$. Figure~\ref{fig6}(b) compares the temperature dependence of the average number of PNIPAM-PNIPAM hydrogen bonding interactions for the suspensions of 16  linear chains, for the network model and for the 48 linear chains with positional restraints for $T>T_{VPT}$. For all systems, at temperatures below $T_{VPT}$, the polymer is characterized on average by one hydrogen bond between PNIPAM hydrophilic groups every $\sim13$ repeating units. On the contrary, for temperatures higher than $T_{VPT}$, only by explicitly considering the polymer aggregation, a net increase of interactions is observed. In particular, in the system of 48 chains the number of hydrogen bonding interactions reaches a value of $\sim0.35$ per repeating unit for temperatures higher than $T_{VPT}$ which is compatible with Fourier transform infrared spectroscopy measurements\cite{MaedaL2000} performed on polymer suspensions at a concentration of 16.7~wt\%. These experiments revealed that in the globule state about 13~wt\% of PNIPAM carbonyl groups is involved in intra- or inter-chain hydrogen bonding interactions suggesting that the affinity between PNIPAM chains increases with concentration.

Another type of interaction that influences the solution behavior of PNIPAM chains is the hydrogen bonding with water. Each repeating unit of PNIPAM is able to form 2 hydrogen bonds with the acceptor carbonyl group and 1 hydrogen bond with the donor amine group therefore in the fully hydrated state 3 hydrogen bonding interactions with water should be present. The temperature dependence of the average number of PNIPAM-water hydrogen bonds for the linear suspensions of 16 chains and 48 chains as well as for the network, normalized to the number of repeating units, is also reported in Fig. \ref{fig6}(b). At low temperatures, on average 2 hydrogen bonds with water are formed in all systems, while at temperature above $T_{VPT}$ only in the suspension of 48 chains the average number of interactions drastically drops to 0.7. The loss of about one polymer-water hydrogen bond per repeating unit is in agreement with the value determined in the volume phase transition of PNIPAM nanogels by UV Resonance Raman scattering experiments \cite{AhmedJPCB2009}. Overall these findings indicate that the simulation model of linear chains with enhanced polymer aggregation is the best suited to describe the experimental results.

\begin{figure*}[htbp]
	\centering
  \includegraphics[width=\textwidth]{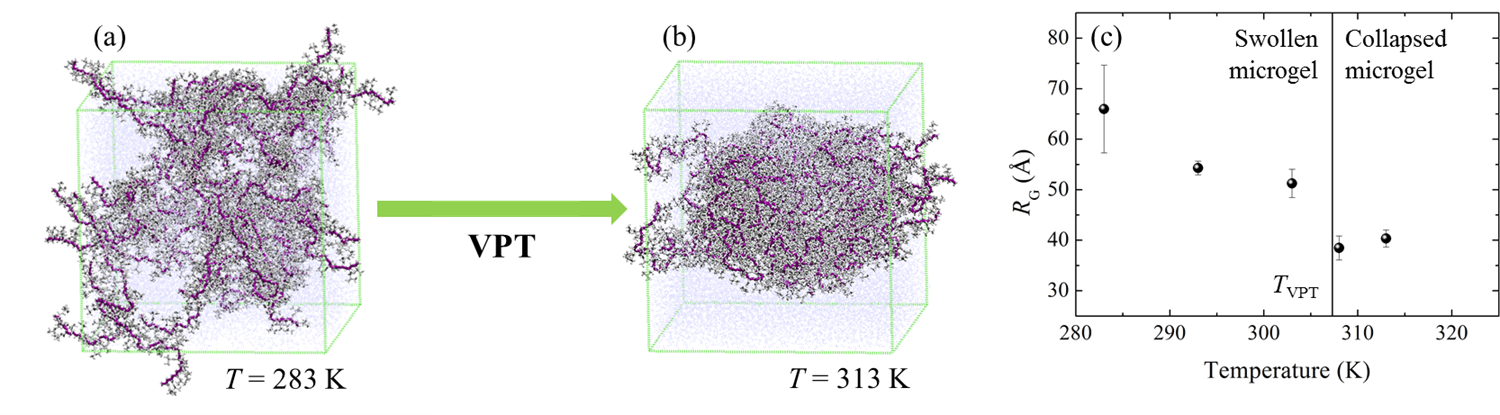}    
  \caption{Model of the microgel network before (a) and after (b) the VPT. The software VMD~\cite{HumphreyJMG1996} was employed for graphical visualization. (c) Calculated radius of gyration $R_g$ across the VPT. The VPT temperature $T_{VPT}$ obtained by EINS is reported with a black line.}
  \label{fig7}
\end{figure*}

By adopting this simulation model, that properly captures the experimental data, we now provide a description of the global structural changes observed at the VPT. Figure~\ref{fig7}(a) and (b) show two representative simulation snapshots at a temperature below and above $T_{VPT}$, which clearly indicate the collective aggregation of polymer chains above $T_{VPT}$. The formation of these aggregated configurations and the evolution of the global structural changes can be monitored by calculating the temperature dependence of the average radius of gyration $R_g$ of the ensemble of polymer chains. This can be obtained as:
\begin{eqnarray}
	R_g=\left(\frac{\sum_{i}\|\textbf{r}_i\|^2m_i}{\sum_{i}m_i}\right)^\frac{1}{2},
	\label{Eq:rg}
\end{eqnarray}
where $m_i$ and $\textbf{r}_i$ are respectively the mass and the position of the $i$-th atom with respect to the center of mass of the polymer chains. As depicted in Figure~\ref{fig7}(c), when temperature is increased, the average radius of gyration slightly decreases, suggesting the presence of small structural alterations, until a sudden reduction occurring at temperatures higher than 303~K. This behavior is can be ascribed to intra- and inter-chain aggregation, that occurs at a transition temperature comparable with that determined from EINS experiments for the microgels suspensions at 30~wt\%.

\section{Discussion and conclusions}
\label{Sec:Discussion}
In this work we have studied the behavior of concentrated PNIPAM microgels suspensions at the atomic level by combining elastic incoherent neutron scattering experiments to all-atom molecular dynamics simulations and we compared the derived information to those obtained at a macroscopic length scale by DSC measurements. Our investigation extends the characterization of the dynamical properties of microgels to previously unexplored concentrations. Indeed, prior to this work, both linear chains and network architectures were examined for samples with a PNIPAM mass fraction limited to about 30~wt\%~\cite{OsakaJCP2009,RetamaMMOL2008,BischofbergerSR2015}. The present contribution complements these works through a characterization of microgels suspensions with a polymer content between 30~wt\% and 95~wt\%.
Due to the high concentration of these systems, that are macroscopically arrested, it is important to check whether the polymeric degrees of freedom can be considered in equilibrium on the probed time-scales for the examined range of temperatures. To this aim, we have repeated EINS experiments in two successive runs, showing that the occurrence of the VPT and the observed  behavior of dynamical properties can be fully reproduced (see Fig.~\ref{fig5}). Thus, it is legitimate to investigate such high concentration samples around the VPT at the atomic scale.

One of the motivations of the present work was to establish whether the VPT shows a clear signature also in very concentrated suspensions. Indeed, as discussed in Sec. \ref{sec:intro}, the swelling-deswelling of the microgels originates from the underlying coil-to-globule transition of linear chains. Such phenomenon can be thus traced back to the behavior of individual macromolecules, either chains or microgels. Hence, in the vast majority of studies, this issue is tackled through the investigation of very dilute samples. In this work, we adopt a different perspective and speculate whether an echo of the behavior at the single-particle level remains visible also for very concentrated systems. Since at molecular level the transition is caused by a reduced hydrophilicity of the polymer at high temperatures, at collective level the VPT should manifest as the formation of aggregates between individual macromolecules. Such a phenomenology has been discussed to some extent in the pioneering work of Bischofberger and Trappe~\cite{BischofbergerSR2015}. In addition, the collective character of the VPT transition was discussed in the rheological study by Howe and coworkers~\cite{HoweACIS2009}. The present study offers a complementary view to these works by addressing the molecular mechanisms and atomic dynamics taking place at the VPT.
For samples up to about 70~wt\%, our EINS measurements clearly reveal that a signature occurs in the dynamics of the PNIPAM hydrogen atoms in correspondence to the VPT. Such signature appears as a sudden drop of the MSD, signaling a slow-down of the dynamics taking place at $T_{VPT}$, which is less and less pronounced with decreasing water content. The slow-down is the result of the occurrence of an aggregation process between individual microgels (or chains), forming an attractive gel state. When the PNIPAM content is increased, the rapid decrease of the MSD gets smoother due to the overall slowing down  of polymer dynamics even below room temperature.

The collective signature of the VPT is also highlighted by the results of our MD simulations. Indeed, to detect an MSD decrease comparable to that observed in experiments,  we should be able to reproduce a gel transition, a task that is too ambitious at the atomistic level. Thus, to capture the essential features of this phenomenon, we use a model of several individual chains that we constrain to remain in an aggregated structure for $T>T_{VPT}$. The failure of unconstrained simulations to reproduce the jump in the MSD can be attributed  to different effects. On one hand, the polymer network model, while faithful in reproducing the dynamics in the interior of a microgel, turns out to be unsuitable to describe an aggregation phenomenon because of its infinite connectivity which prevents it from fully collapsing. For the linear chains, one important aspect to consider is the small size of the simulation model (roughly 10~nm in box side) which may not be able to cover the necessary range of length-scales to be probed within the experiments. In this small system, at full aggregation of the chains the surface-to-bulk ratio of PNIPAM atoms is still relatively large, so that we do observe a small decrease of the unconstrained MSD (see Fig~\ref{fig6}(a)) but not enough to reproduce the experimental values. Equilibration issues could further arise in the high $T$ simulations, due to the increase of the hydrogen bond lifetime (not shown). Finally, we notice that recent simulation studies have shown that it is necessary to re-parameterize force fields upon changing dielectric environment~\cite{DalgicdirJPCB2019}. Such situation could become relevant $T>T_{VPT}$, where the collapse of the polymer may have significant effects. We leave the investigation of this issue to future work.

Having listed all possible sources of discrepancy, it is nonetheless fair to say that the simulations with enhanced aggregation are in quantitative agreement with experiments, confirming the working hypothesis that microgels connect into an attractive state, and this feature is present in the atomic dynamics of concentrated systems, in the same way as it is in the mesoscopic dynamics of individual particles. We also notice that the agreement between experiments and simulations, without adjustable parameters, suggests that, at the high concentrations probed in these systems, the underlying polymer architecture is not relevant, since a similar behavior is found for (measured) microgels and (simulated) polymer chains. 

Another interesting result is the good correspondence between the microscopic estimates of $T_{VPT}$ obtained from EINS and the corresponding macroscopic ones from calorimetry. This confirms again the occurrence of the transition across all scales and points to the intriguing finding that there exists an optimal PNIPAM concentration, roughly  $c\sim43$~wt\%, where $T_{VPT}$ displays a minimum. This value is found to be roughly the same as the one above which water crystallization does not occur in any range of temperatures~\cite{ZanattaSA2018}. We thus argue that a water concentration of 57~wt\% corresponds to the ideal hydration of the microgel, a situation where all water molecules are within the microgel solvation shell and are influenced by the presence of the polymer. Overall these findings suggest that the solvent energetics, i.e. the difference between the energy of bulk and shell water, plays a major role in controlling the phase behavior of an amphiphilic polymer~\cite{BischofbergerSR2014}.

Finally, it is important to remark that the volume phase transition in PNIPAM-based systems originates from the complex amphiphilic nature of the polymer that is composed by a hydrophobic backbone decorated with side chain groups containing both hydrophilic amide moieties and hydrophobic isopropyl ones. DSC measurements show that the VPT occurs with a positive transition enthalpy (Fig.~\ref{fig1}), which is consistent with the molecular mechanism provided by the MD simulations (Fig.~\ref{fig6}(b)) where a significant reduction of PNIPAM-water interactions from the swollen to the collapsed state of the polymer network is not compensated by the formation of an equivalent number of PNIPAM-PNIPAM interactions. Figure \ref{fig6}(b) also indicates that residual PNIPAM-water hydrogen bonds are formed even at $T\gg T_{VPT}$. In agreement with experimental findings, which suggest that in the phase-separated region the residual concentration of PNIPAM ranges from $c\sim40$~wt\% to 70~wt\%~\cite{WuPRL1998,FernandezPRE2002}, we thus provide evidence that, even well-above the VPT temperature, the polymer does not become fully hydrophobic and a relatively large amount of water still remains within the microgel~\cite{PeltonJCIS2010}.

\begin{acknowledgments}
We acknowledge ILL for beamtime and CINECA-ISCRA for computer time. LT, MB, EC and EZ acknowledge support from European  Research Council (ERC-CoG-2015, Grant No. 681597 MIMIC); LT, EB, EC and EZ from MIUR (FARE project R16XLE2X3L SOFTART) and from Regione Lazio, through L.R. 13/08 (Progetto Gruppo di Ricerca GELARTE, n.prot.85-2017-15290).
\end{acknowledgments}

\section*{AIP Publishing Data Sharing Policy}
The data that support the findings of this study are available from the corresponding author upon reasonable request.

\end{document}